# Microstructure of strained $La_2CuO_{4+\delta}$ thin films on varied substrates

Jiaqing He and Robert F. Klie

Center for Functional Nanomaterials, Brookhaven National Laboratory, Upton, NY 11973, USA

Gennady Logvenov and Ivan Bozovic

Condensed Matter Physics and Materials Science Department, Brookhaven National Laboratory, Upton, NY 11973, USA

Yimei Zhu[a]

Center for Functional Nanomaterials, Brookhaven National Laboratory, Upton, NY 11973, USA

## Abstract

Layered perovskite $La_2CuO_{4+\delta}$ (LCO) thin films were epitaxially grown on $SrTiO_3$ (STO) and $LaSrAlO_4$ (LSAO) substrates by atomic-layer-by-layer molecular beam epitaxy. The lattice defects and residual strain in these films were investigated by means of transmission electron microscopy and electron energy loss spectroscopy. The LCO films showed a high epitaxial quality with flat interfaces and top surfaces. Misfit dislocations with Burgers vector a<010> and shear defects were frequently observed at or near the film/substrate interfaces and in the films, respectively, for all the LCO films. In one LCO film, grown on STO at the highest temperature, 700 °C, was a two layered structure. The top layer was identified as rhombohedral $LaCuO_2$ by electron energy loss spectroscopy combined with electron diffraction. In addition, stacking faults were observed in plane-view micrographs of one LCO film grown on the STO substrate. The residual strains were evaluated from the spacing of misfit dislocations at the film/substrate interface and from the split electron diffraction patterns. Possible mechanisms of strain relaxations are discussed based on the observed defects.

---

[a] electronic mail: zhu@bnl.gov

# 1. Introduction

The epitaxial growth of $La_2CuO_{4+\delta}$ (LCO) superconductor films has attracted much attention because of the remarkable effects of epitaxial strain on the critical temperature, $T_c$ [1-15]. Basically, substrates provide two types of in-plane strain for the films: compressive strain if the lattice constant of the film, $a_f$, is greater than the lattice constant of the substrate, $a_s$, and tensile strain if $a_f < a_s$ (Figure 1). Usually, in-plane compressive strain raises $T_c$, while in-plane tensile strain lowers it. Indeed, what is relevant is the residual strain that strongly depends on the lattice mismatch between the film and the substrate. The state of strain in a film can differ considerably from that found in bulk materials with respect to the configuration of microstructure and lattice defects. Thus, the electrical properties (e.g., $T_c$) of a strained film also can differ from those in the bulk. The lattice mismatch between a thin film and a substrate is one of the important parameters that greatly influence on the film's structure, morphology, and its remaining strain. The energy stored within an epitaxial film due to the lattice misfit strain increases with the film's thickness. At a certain critical thickness, $t_{cr}$, it is energetically favorable to relieve the misfit strain by formation of dislocations at or near the substrate/film interface. Usually, the configuration of the defect and the residual strain depend not only on the lattice mismatch and the film thickness but also on the nature of substrate and on the processing conditions.

At room temperature, LCO is orthorhombic crystal (Cmca) with lattice parameters $a_0 = 0.53346$ nm, $b_0 = 0.54148$ nm, and $c_0 = 1.31172$ nm [16]. LaSrAlO$_4$ (LSAO) has a tetragonal structure with space group I4/mmm and lattice parameters $a = 0.3755$ nm and $c = 1.262$ nm [17]. SrTiO$_3$ (STO) exhibits a typical cubic perovskite structure (Pm$\bar{3}$m) with a lattice parameter $a = 0.3905$ nm. While the structure of LCO is similar to that of LSAO, one difference lies in the small ortho-rhombic distortion (about 0.5%) of the LCO lattice. For convenience in comparison with LSAO and STO, in the following discussion, we regard the LCO lattice as a pseudo-tetragonal one with lattice parameters $a = (a_0^2 + b_0^2)^{1/2}/2 = 0.38005$ nm and $c = c_0 = 1.31172$ nm. The lattice mismatch is by given by

$$f = \frac{a_f - a_s}{a_f} \qquad (1)$$

where $a_f$ and $a_s$ are the lattice parameter of the film and the substrate, respectively. From the lattice parameters listed above, the lattice mismatch is +2% for LCO /LSAO and -2.75% for LCO /STO. The mismatch with positive sign can induce an in-plane compressive stress in the film while that with negative sign produces an in-plane tensile stress.

In this paper, we report investigation of the microstructure and remaining strains in LCO films grown on STO and LSAO substrates, by means of transmission electron microscopy (TEM) combined with electron energy loss spectroscopy. We focus on the possible effects of the substrate, microstructure and defect configuration on the level of strain in the films.

## 2. Experiments

The films used in this study were all grown using the atomic-layer-by-layer molecular beam epitaxy (ALL-MBE) technique, which has been described in detail previously [18,19]. LCO films were synthesized in a unique multi-chamber molecular beam epitaxy system, designed to enable atomic-layer engineering and optimization of complex oxide materials. The growth chamber contains 16 metal sources and a source of pure ozone. Deposition rates are mapped using a scanning quartz-crystal oscillator, and they are controlled accurately in real time by a 16-channel atomic absorption spectroscopy system. The film's growth is monitored by a reflection high-energy electron diffraction (RHEED) system and other state-of-the art surface analysis tools [19]. The films were also characterized in detail by atomic-force microscopy (AFM) and x-ray diffraction (XRD). Here, we describe the microstructure of three samples: A is a 50- nm-thick LCO film deposited on LSAO at $T_s \approx 650\ ^0C$, samples B and C are 30-nm-thick LCO films on STO deposited at $T_s \approx 650\ ^0C$ and $T_s \approx 700\ ^0C$, respectively. TEM and HRTEM investigations were carried out in JEOL -3000F and -4000EX microscopes. The compositional homogeneity of the LCO thin film was investigated by electron energy loss spectroscopy (EELS). High-resolution image numerical simulations were carried out with the MacTempas computer program with the following parameters as input: spherical aberration of 1mm, defocus spread of 8 nm, semiconvergence angle of illumination of 0.55mrad, and diameter of the objective lens aperture of 7 nm.

## 3. Results

Low-magnification lattice images of samples A, B, and C, taken with the incident electron beam parallel to the [110] zone axis of the films, are shown in Figs.2 (a)–(c), respectively. Very flat top surfaces and interfaces are seen in all the films. The horizontal arrowheads in the images denote the interfaces between the thin films and the substrates. The epitaxial nature of the 50-nm-thick LCO film for sample A and 30-nm-thick films for samples B and C are evident from Fig. 2. The films A and B are homogeneous, while the film C has two layers, with the $LaCuO_2$ layer on top, as discussed below.

Figure 3 shows selected-area electron diffraction (SAED) patterns for three samples along the [110] LCO zone axis acquired with an aperture that simultaneously collects scattered electrons coming from the LCO films, and from the LSAO or STO substrates. The orientation relationships between the LCO film and the LSAO substrate can be described as $[001]_{LCO} \parallel [001]_{LSAO}$ and $[1\bar{1}0]_{LCO} \parallel [010]_{LSAO}$, as shown in Fig. 3(a). The SAED pattern from the LCO/STO film in Figs. 3(b) and (c) clearly demonstrate similar orientation relationships, viz. $[001]_{LCO} \parallel [001]_{STO}$ and $[1\bar{1}0]_{LCO} \parallel [010]_{STO}$. In Fig. 3, some distinct splits of diffracted spots also can be detected along both the out-of-the-plane and the in-plane direction. Image calibration allows us to determine the lattice parameters of the LCO films and top layer in sample C (given in Table 1), assuming a tetragonal LSAO lattice with $a = 0.3754$ nm and $c = 1.2635$ nm, see Fig. 3 (a), and the cubic STO lattice with $a_{STO} = 0.3905$ nm, see Fig.3 (b)-(c).

The lattice parameters of the three samples, A, B and C in Table 1 indicate that massive relaxation mechanisms exist within all of them although some residual strains remain in the LCO films that are under in-plane compressive strain (in the sample A) or under in-plane tensile strain (in samples B and C). Such a lattice misfit relaxation is accommodated through several mechanisms, i.e., via formation of: (i) a network of misfit dislocations at or near the interfaces (Fig.4), often considered as the commonest mechanism; (ii) planar defects such as shear defects (Fig. 5) and stacking faults (Fig. 7), and, (iii) possible secondary phase $LaCuO_2$ (Fig. 6).

It is well known that lattice mismatch can be accommodated either by straining the lattice or by the formation of misfit dislocations at the interface depending on the elastic properties of the materials and the film's thickness. To explore this accommodation, the interfaces of all the films were investigated by HRTEM. For the LCO film on the LSAO substrate, the film-substrate interface appears atomically sharp; over large areas only a few dislocations can be detected at or close to the LCO/LSAO interface due to small mismatch between LCO film and LSAO substrate. Figure 4 (a) shows the [110] lattice fringe image of the LCO film on the LSAO. One misfit dislocation with a Burgers vector a<010> in the film exhibits a stand-off of 5 nm from the interface between the LCO film and LSAO substrate due to the different elastic properties between LCO film and LSAO substrate [20] or to local composition fluctuation [18,19]. For LCO films on STO, numerous dislocations are observed at the LCO/STO interface. Figure 4(b) shows a lattice image of an LCO film on STO (sample B). There are two observable misfit dislocations (denoted by vertical arrows at the interface of the LCO film and the STO substrate) with Burgers vector of a[010], according to the Burgers circuit analysis surrounding the dislocation core (see Fig.4).

The possible residual mismatch strain in LCO/STO film was assessed by consideration of the separation distance and the Burgers vector of these misfit dislocations. As measured over a long distance along the interface we determined an averaged value of 14.6 nm for the separation distance between two dislocations with the Burgers vector a<100>. Assuming that the misfit strain in the film is completely released by generation of misfit dislocations, the separation distance S of the misfit dislocations can be calculated from the following equation:

$$S = b/f = bd_f / (d_s - d_f) \qquad (2)$$

where b is the magnitude of the Burgers vector of misfit dislocations in the interface plane, while $d_f$ and $d_s$ are the bulk values of inter-planar distances of the film material and the substrate, respectively. Inserting the value of the Burgers vector b = 0.39 nm, and using $d_f$ = 0.38 nm, $d_s$ =

0.39 nm, we obtained a dislocation spacing of 14.2 nm. The experimental value, 14.6 nm, is therefore almost equal to the value expected for a completely relaxed system. This result, corresponding to that in Fig. 2(c), suggests that the mismatch strain is almost completely relieved in the film.

Figure 5 (a) shows one type of planar defects, an octahedral edge-shear defect, in an LCO film that originates at the interface between the LCO film and the STO substrate due to non-stoichiometry. The two domains are translated by 1/6 of the unit cell dimension along the (031) plane, **R**= 1/6[031]$_{LCO}$. Figure 5 (b) is a possible model for such a shear defect, showing an excess of oxygen atoms and possibly also of La cations. The simulation high-resolution image of this modeling is inserted in Figure 5(a); it revealed a good fit with the lattice images. The possible mechanisms of formation of shear defects are lattice mismatch strain [23], steps [15], or stacking faults [12]. However, neither steps nor stacking faults are seen in Fig. 5(a), and therefore, the shear defects also may contribute to the relaxation of the misfit strain.

Figure 6 (a) is a cross-section image of a LCO film, taken along the [100] zone axis of the STO substrate that clearly shows two different structural layers distribution. The interface between the two layers is quite rough due to the variation in thickness of the top layer (from 10 nm to 25 nm away from the LCO/STO interface), however, the top surface is flat. To determine the film's chemical composition, a line scan of EELS measurement was performed from the STO substrate, via the LCO film and the secondary-phase layer, to the glue layer, as marked by A-B in Fig. 6 (a) [24]. The energy range from 450 eV to 1000 eV was monitored, which included the O K-, La M$_{4,5}$-, and Cu L$_{2,3}$-edges. Figure 6 (b) shows the EELS spectra obtained from these four typical ar-eas (substrate, film, secondary phase and glue) in the scanning transmission electron microscopy (STEM) mode with a probe size 1 nm in diameter. In these spectra, the energy resolution is about 1.2 eV for the FWHM (full width at half maximum) of the zero-loss peak. We note that the intensity of the La peak in spectra 3 is much less than that in spectra 2, indicating an La deficiency in the secondary phase. Figure 6 (c) presents the line scanning EELS results after background subtraction and removal of the thickness effect. It reveals show how the composition of the sample (the La-, Cu- and O- content) varies with scanning position. Clearly, the top layer still contains all three elements, La, Cu and O, but the amount of La is about half of that in the LCO film. Taking into account the lattice parameter obtained from the electron diffraction pattern in Fig. 6 (c), we conclude that the top layer is $LaCuO_2$, i.e., rhombohedral with a group space of $R\bar{3}m$ and the lattice parameters a = b = 0.38 nm and c = 1.70 nm.

Figure 7 shows plane-view micrographs of a wormlike meandering defect in the LCO film on the STO substrate. Two enlarged lattice images of the marked areas, taken along the [001] direction, indicate that the defect is a stacking fault rather than a misfit dislocation. The surface of STO substrates often shows steps, terraces and kinks, due to miscuts, that strongly influence on the microstructure of the LCO films. Hence, these wormlike stacking faults are most likely formed as the consequence of the specific surface structure of STO and the growth mode of the LCO films.

## 4. Discussion

The above results show that the quality and perfection of LCO films strongly depend on the substrates with different lattice mismatch to the film material. The effects of substrate extend not only to generation of misfit dislocations but also of planar defects and formation of secondary phase(s). For a system that exhibits a lattice mismatch, one can define [25] the film's critical thickness, $t_{cr}$, can be defined as

$$t_{cr} = \frac{(1-\nu\cos^2\theta)b}{8\pi f(1+\nu)}\ln(\frac{\beta T_{cr}}{b}) \quad (3)$$

Here, b is the magnitude of the Burgers vector of the misfit dislocation in terms of the substrate lattice, θ is the angle between the Burgers vector and the dislocation line of the misfit dislocation, f is the misfit, ν is the average Poisson ratio of the film and the substrate, and β is the cut-off radius of the dislocation core with β≈4 [26]. To calculate the critical thickness in *c*-axis-oriented single-crystal films, the misfit segments of half-loops were assumed to be edge dislocations with Burgers vectors of a[010] type, gliding on the {010} plane. Using the parameters b = 0.3905 nm, f = 0.0275, ν = 0.3 and θ = 90$^0$, we calculated the critical thickness for STO substrate to be $t_{cr} \approx 1.02$ nm.

If the thickness of the film is less than the critical thickness $t_{cr}$, the mismatch between the substrate and the film is accommodated by elastic lattice strain, i.e., no crystal defects such as misfit dislocations will be formed. When the film's thickness exceeds $t_{cr}$, the formation of misfit dislocations is energetically more favorable than the elastic strain accommodation [25]. The critical thickness depends on the level of lattice mismatch and on the elastic properties of the two materials. Here we found the strain to be almost completely relaxed by forming defects, since all three films are much thicker than the $t_{cr}$. We also note that the superconducting transition (critical) temperature was close to the *bulk's* value. The high perfection of the LCO film on LSAO is based on the very small lattice mismatch in this system; nevertheless some misfit dislocations were found here as well at or near the interface because the critical thickness for LCO films on LSAO is smaller than the film thickness, d = 30 nm. According to the measurements of spacing and Burgers vectors of the misfit dislocations, we also would expect no residual strain in LCO films on STO.

We noted that in sample C there was a $LaCuO_2$ secondary phase in LCO film. Similar observation ($LaCuO_{3-\delta}$ at the interface between LCO films and STO substrate) was reported by Houben [15]. Kong et al. [11] also observed a secondary phase at the interface between a $La_2CuO_4F_x$ film and an STO substrate. In those studies, precipitates were detected close to the interface between the film and the substrate. In the present study, we have observed the secondary phase $LaCuO_2$ at the top surface of LCO films. The formation of $LaCuO_2$ precipitates in the film matrix could originate from local fluctuations in chemical composition and possibly from different surface mobility of constituent atoms. Since the lattice parameters of $LaCuO_2$ are different from those of LCO, the formation of precipitates also may contribute to partially relieving the misfit strain that is incompletely accommodated by misfit dislocations.

## 5. Conclusions

The microstructure and residual strain of LCO films grown on STO and LSAO substrates by ALL-MBE technique were investigated at atomic scale by means of TEM, combined with HRTEM, electron diffraction and EELS analysis. The strain in LCO films on both types of substrates was found to be mostly relaxed by the formation of misfit dislocation cores, although the small lattice mismatch was accommodated by elastic strain in the LCO film on the LSAO substrate. In all the films, we observed planar defects (shear defects and stacking faults) as well as one type of misfit dislocations (with the Burgers vectors a<010>) in the interface area between the film and the substrate. A secondary-phase formation was also facilitated strain relaxation.

## Acknowledgment

This work was supported by the U.S. Department of Energy, Division of Materials, Office of Basic Energy Science, under Contract No. DE-AC02-98CH10886.

**Table 1**

The lattice parameters of LCO films on LSAO and STO substrates determined from super-imposed electron diffraction patterns with reference to the reflection spots of LSAO or STO.

| Sample | Growth temperature (°C) | substrate | Film thickness (nm) | Parameter | |
|---|---|---|---|---|---|
| | | | | $a_{LCO}$ (nm) | $c_{LCO}$ (nm) |
| Bulk | – | – | – | 0.38005 | 1.31172 |
| A | 650 | LSAO | 50 | 0.3798 | 1.3265 |
| B | 650 | STO | 30 | 0.3809 | 1.3111 |
| C | 700 | STO | 30 | 0.3801 | 1.3107 |

# Figure Captions

Fig. 1. Two different types of in-plane strain in LCO films on various substrates.

Fig. 2 a-c. Low-magnification cross-section TEM images of samples A, B and C, respectively. In all the samples, there are sharp and flat top surfaces and interfaces between films and substrates are observed. A secondary phase is visible close to the top surface in sample C.

Fig. 3. SAED patterns taken along (a) the [110] LSAO and (b) and (c) the [100] STO zone axes. They result from simultaneous diffraction coming from the films and from parts of the substrates. Some distinct splits spots are detected along both the out-of-plane and the in-plane directions in (a) and (b); besides these, some extra diffraction spots are observed in (c) and its partially enlarged image (bottom right).

Fig. 4. Enlarged lattice images showing the sharp interfaces between films and substrates. (a) A misfit dislocation near the interface between LCO and LSAO; (b) two misfit dislocations with Burgers vectors a[010] at the interface of LCO and STO. The positions of dislocations are marked by the arrows.

Fig. 5. (a) An octahedral corner-shearing defect with displacement vector R = 1/6[031]; and (b) a possible model of the shear defect, with the triangle symbols representing La cation vacancies. The simulated lattice image of this model is shown in insert (a)

Fig.6. (a) Cross-section lattice images of an LCO film on STO substrate. A secondary phase with structure different from LCO is delineated by the dotted curve lines. (b) EELS spectra in the energy range from 450 eV to 950 eV obtained from the four typical areas in Fig. 6(a). (c) The compositional distribution (La, Cu and O) obtained from the line scan of EELS spectra from the regions of substrate STO, the LCO film, the secondary phase (S) and the glue in Fig. 6(a). The horizontal axis is the distance (in nm) from the interface between LCO film and STO substrate.

Fig.7. A worm-like meandering planar defect from plane-view low-magnification image of an LCO film on STO with two enlarged lattice images (circled areas) showing lattice mismatch due to stacking faults.

# Figure 1

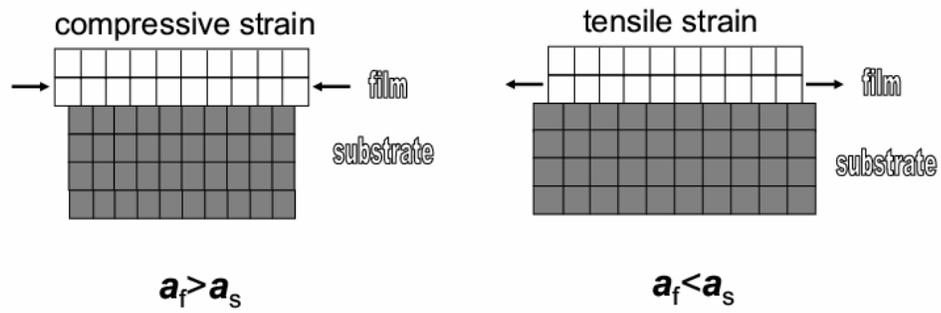

figure 2

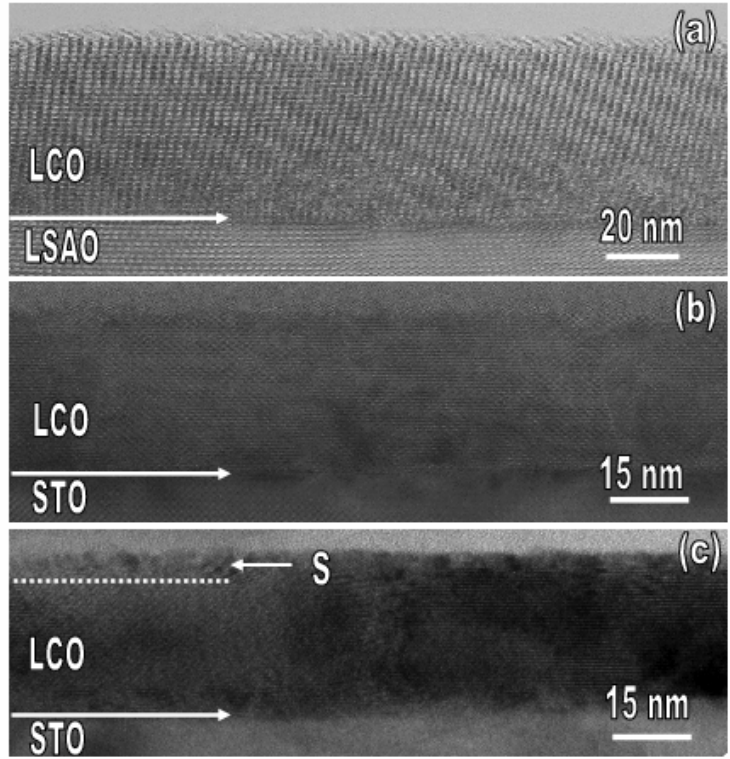

figure 3

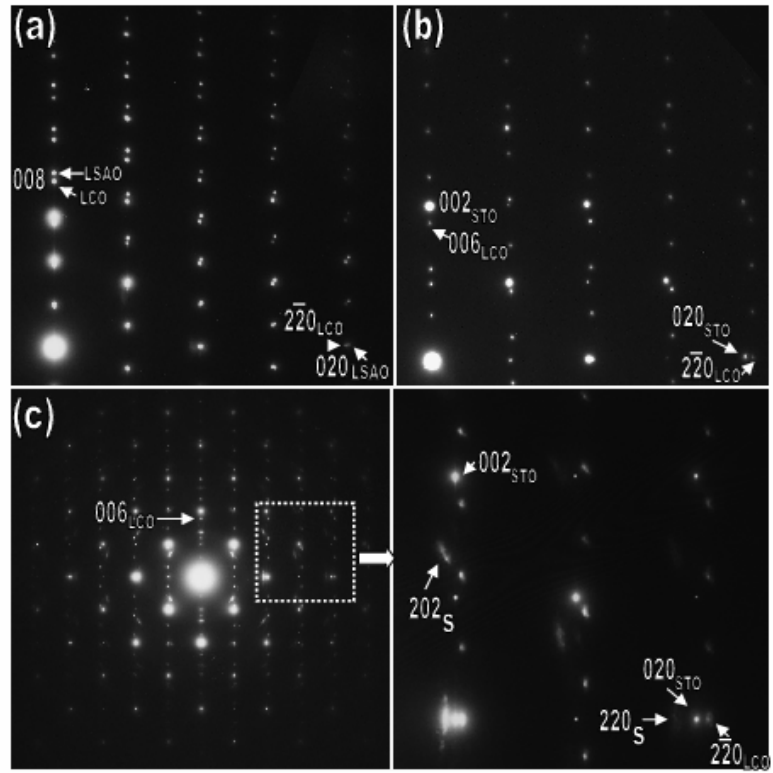

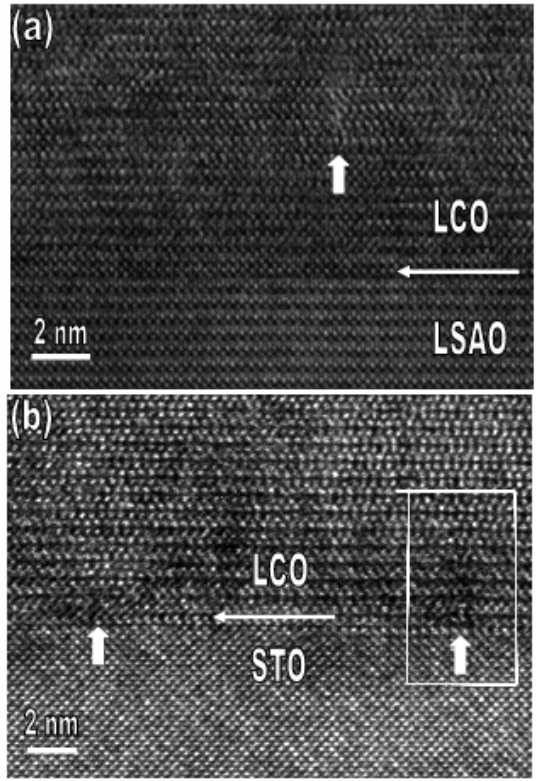

figure 4

figure 5

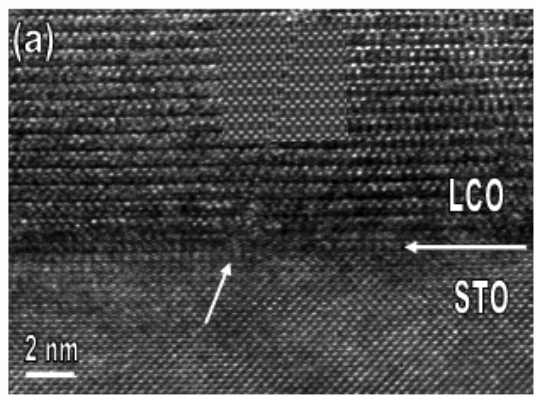

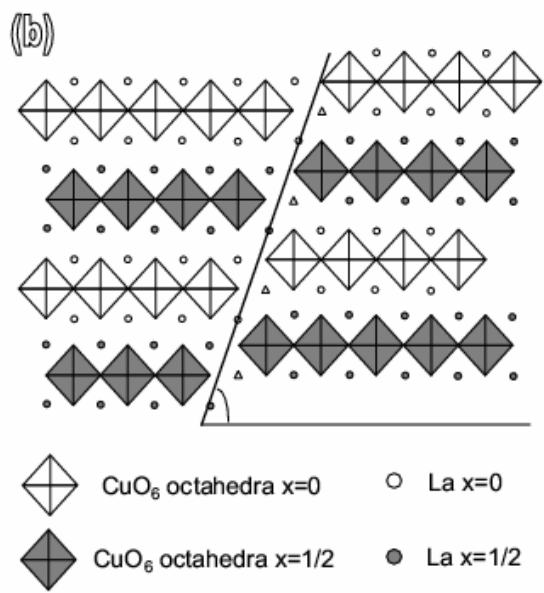

| CuO$_6$ octahedra x=0 | ○ La x=0 |
| CuO$_6$ octahedra x=1/2 | ● La x=1/2 |

figure 6

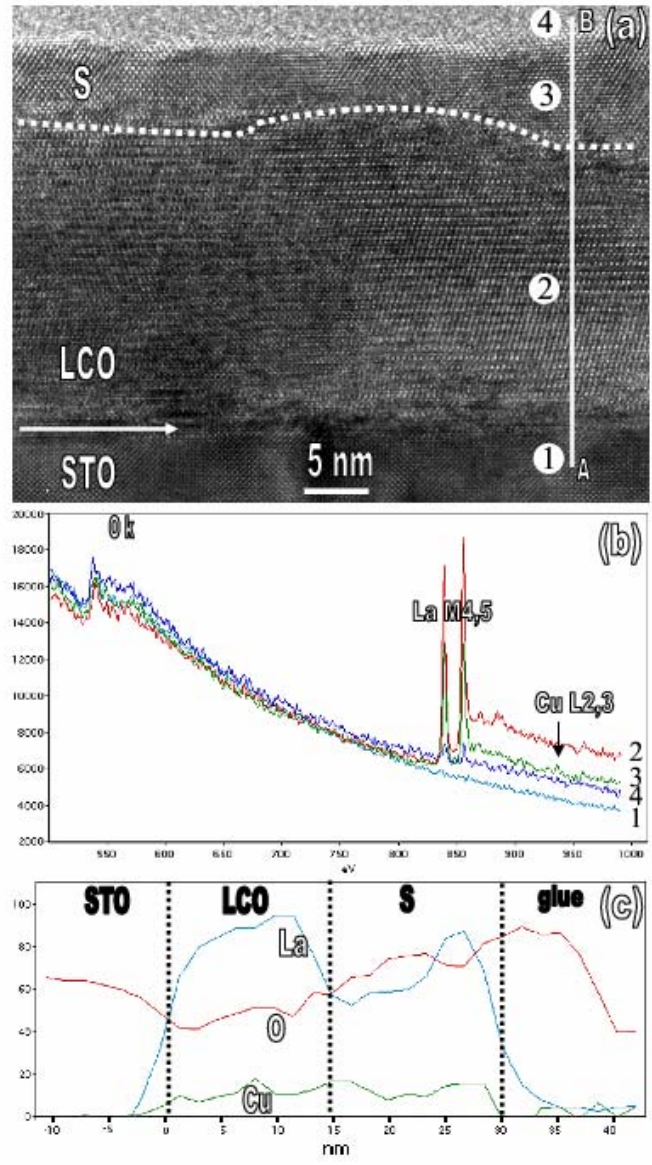

figure 7

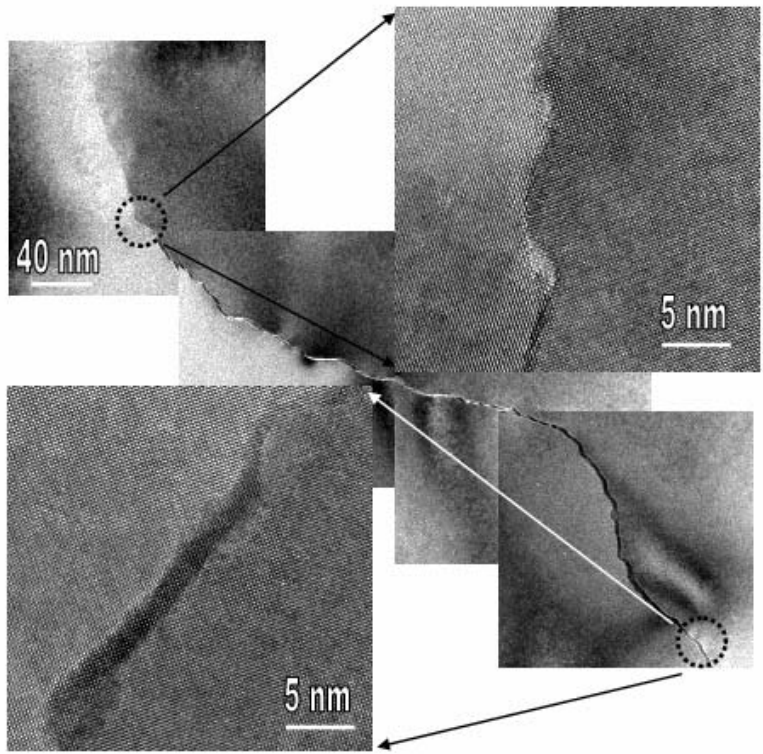